\documentclass[twocolumn,doublespacing,showpacs,preprintnumbers,amsmath,amssymb]{revtex4-1}
\usepackage{amssymb}
\usepackage{txfonts}
\usepackage{graphicx}
\usepackage{subfigure}
\usepackage{dcolumn}
\usepackage{color}
\usepackage{bm}
\usepackage{sidecap}
\usepackage[colorlinks,linkcolor=blue,anchorcolor=blue,citecolor=blue]{hyperref}

\begin{document}

\title{Long-Distance Continuous-Variable Quantum Key Distribution over $202.81$~{\rm km} of Fiber}

\author{Yichen Zhang$^{1}$}
\author{Ziyang Chen$^{2}$}
\author{Stefano Pirandola$^{3}$}
\author{Xiangyu Wang$^{1}$}
\author{Chao Zhou$^{1}$}
\author{Binjie Chu$^{1}$}
\author{Yijia Zhao$^{1}$}
\author{Bingjie Xu$^{4}$}
\author{Song Yu$^{1}$}
\thanks{yusong@bupt.edu.cn}
\author{Hong Guo$^{2}$}
\thanks{hongguo@pku.edu.cn}

\affiliation{$^1$State Key Laboratory of Information Photonics and Optical Communications, Beijing University of Posts and Telecommunications, Beijing 100876, China }

\affiliation{$^2$State Key Laboratory of Advanced Optical Communication Systems and Networks, Department of Electronics, and Center for Quantum Information Technology, Peking University, Beijing 100871, China}

\affiliation{$^3$Computer Science and York Centre for Quantum Technologies, University of York, York YO10 5GH, United Kingdom}

\affiliation{$^4$Science and Technology on Security Communication Laboratory, Institute of Southwestern Communication, Chengdu 610041, China}

\date{\today}

\begin{abstract}
Quantum key distribution provides secure keys resistant to code-breaking quantum computers. The continuous-variable version of quantum key distribution offers the advantages of higher secret key rates in metropolitan areas, as well as the use of standard telecom components that can operate at room temperature. However, the transmission distance of these systems (compared with discrete-variable systems) are currently limited and considered unsuitable for long-distance distribution. Herein, we report the experimental results of long distance continuous-variable quantum key distribution over $202.81$ {\rm km} of ultralow-loss optical fiber by suitably controlling the excess noise and employing highly efficient reconciliation procedures. This record-breaking implementation of the continuous-variable quantum key distribution doubles the previous distance record and shows the road for long-distance and large-scale secure quantum key distribution using room-temperature standard telecom components.
\end{abstract}

\maketitle



\emph{Introduction.} $-$ The BB84 protocol~\cite{BB84_1984} started the era of quantum key distribution (QKD)~\cite{BB84_1984,Ekert_PhysRevLett_1991,Pirandola_RevModPhys_2019,Feihu_RevModPhys_2019}, providing a way to securely generate secret keys between two remote users by exploiting the laws of quantum mechanics. Combining this method with one-time pad provides ultimate physical-layer protection to the transmission of confidential messages. In general, for cost-effective implementations, QKD protocols are formulated in a prepare-and-measure fashion, where classical information is encoded in nonorthogonal quantum states: these are randomly prepared by Alice (the sender) and then transmitted to Bob (the receiver) through an insecure quantum channel. At the output of the channel, the states will be measured by Bob, so as to retrieve the encoded classical information. Depending on the setting, this measurement may consist of single-photon detections or coherent measurements, such as homodyne or heterodyne detections. The latter are certainly more attractive for commercial deployment, due to their room-temperature operation and compatibility with the current telecommunication infrastructure~\cite{Zhang_NaturePhoton_2019,Zhang_QuantumSciTechnol_2019,Karinou_PhotonTechnolLett_2018,Eriksson_CommunicationsPhysics_2019}. Protocols that exploit such coherent measurements and encode classical information by modulating states of
an optical mode are today very popular and known as continuous-variable QKD (CV-QKD) protocols~\cite{Pirandola_RevModPhys_2019,Weedbrook_RevModPhys_2012,Lam_NaturePhoton_2013,Usenko_Entropy_2016}.

The most known CV-QKD protocol is the seminal GG02 protocol: this is based on the Gaussian modulation of the amplitudes of coherent states and homodyne detection of the channel output~\cite{Grosshans_PhysRevLett_2002,Grosshans_Nature_2003}. This protocol later evolved into various other Gaussian protocols~\cite{Weedbrook_PhysRevLett_2004,Patron_PhysRevLett_2009,Pirandola_NaturePhys_2008} and it has been the subject of increasingly-refined security proofs~\cite{Acin_PhysRevLett_2006,Patron_PhysRevLett_2006,Furrer_PhysRevLett_2012,Leverrier_PhysRevLett_2015}.
Many experimental demonstrations of GG02~\cite{Jouguet_NaturePhoton_2013,Huang_ScientificReports_2016} and other CV-QKD protocols have been achieved so far (e.g., see~\cite[Sec.~VIII]{Pirandola_RevModPhys_2019} for an overview). The longest distance achieved in CV-QKD is currently $100$~{\rm km} in fiber, with a secure key rate of the order of $50$~{\rm bps}~\cite{Huang_ScientificReports_2016}. Compared to the performance of discrete-variable (DV) QKD protocols, this is a limited transmission distance with a relatively low key rate. 

Here, we report the longest-distance experimental demonstration of CV-QKD, paving the way for closing the gap with the current performance of DV-QKD protocols. In fact, our experiment realizes CV-QKD over the record-breaking distance of about $200$~{\rm km} of fiber channel, doubling the previous record~\cite{Huang_ScientificReports_2016}. More precisely, we achieve the secret key rate of $6.214$~{\rm bps} at a distance of $202.81$ {\rm km} of ultralow-loss optical fiber. We obtain this result thanks to a fully automatic control system and high-precision phase compensation, so that the excess noise can be kept down to reasonably low values. In our experiment, we also use different reconciliation strategies at the various experimental distances considered, with an efficiency of $98$\% for the longest point at $202.81$ {\rm km}.

\begin{figure*}[t]
\centerline{\includegraphics[width=0.92\textwidth]{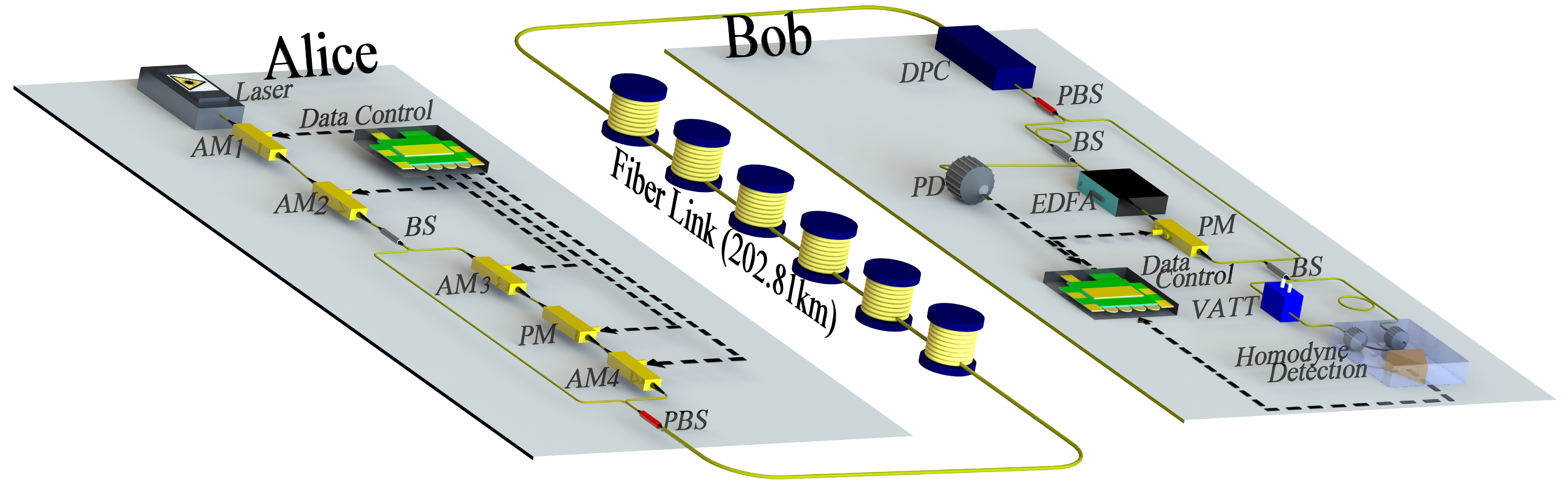}}
\caption{Optical layout of our long-distance CV-QKD system. Alice sends an ensemble of $38$~{\rm ns} weak Gaussian-modulated coherent states to Bob multiplexed with a strong local oscillator (LO) in time and polarization by using a delay line and a polarization beam-combiner, respectively. Then the two optical paths are demultiplexed at Bob's side by a polarizing beam splitter placed after an active dynamic polarization controller. We perform phase modulation on the LO path to select the signal quadrature randomly. Finally, the quantum signal interferes with the LO on a shot-noise-limited balanced pulsed homodyne detector. Laser: continuous-wave laser; AM: amplitude modulator; PM: phase modulator; BS: beam splitter; VATT: variable attenuator; PBS: polarization beam splitter; DPC: dynamic polarization controller; EDFA: Erbium doped fiber amplifier; PD: photodetector.}\label{fig1}
\end{figure*}

\emph{Experimental Setup.} $-$ Our experimental setup is shown in Fig.~\ref{fig1}. At Alice's side, continuous-wave coherent light is generated by a $1550$~{\rm nm} commercial laser diode with a narrow linewidth of $100$~{\rm Hz} (NKT BasiK E15). Two cascaded amplitude modulators (AM$_1$ and AM$_2$, iXblue, MXER-LN-10-PD), each with high optical extinction of $45$~{\rm dB}, generate light pulses at a repetition rate of $5$~{\rm MHz}. 
A very unbalanced $1/99$ beam splitter (General Photonics) divides the pulses into strong local oscillator (LO) pulses and weak signal pulses. The latter are modulated by an amplitude modulator (AM$_3$, Optilab, IM-1550-12-PM) with Rayleigh distribution and a phase modulator (PM$_1$, EOSPACE, PM-OSES-10-PFA-PFA) with uniform distribution to get the zero-centered Gaussian distributions. For security reasons, the signal pulses are then attenuated to a several-photon level by using another amplitude modulator (AM$_4$, iXblue, MXER-LN-10-PD). Finally, the signal pulses are recombined with the LO pulses in a polarizing beam splitter and sent to Bob, each with a duration period of $38$~{\rm ns}.

At the output of the fiber link, signal and LO pulses are demultiplexed by another polarizing beam splitter which is placed after an active dynamic polarization controller whose aim is to optimize the outputs. On the LO path, a customized Erbium doped fiber amplifier consisting of ultra narrow band optical transmission filter (about $0.5$~{\rm GHz} @ $1550.12$~{\rm nm}) is employed for amplifying the copropagated and decreased-power LO to a magnitude that is large enough to amplify the weak quantum signal. A phase modulator (PM$_2$, EOSPACE, PM-5K4-10-PFA-PFA-UV) on the LO path selects the signal quadrature components randomly. The signal path is randomly switched off to do the real time shot-noise unit calibration~\cite{YC_PhysRevApplied_2020}. The signal pulses interfere with the LO pulses on a self-developed shot-noise-limited balanced pulsed homodyne detector whose output is proportional to the signal and LO intensity.

For the fiber link, we have used an ultralow-loss ITU-T G.$652$ standard compliant fiber (Corning\textsuperscript{\textregistered} SMF-28\textsuperscript{\textregistered} ultralow-loss fiber)~\cite{Corning}. The average fiber attenuation (without splices) is $0.16$~{\rm dB/km} at $1550$~{\rm nm}. Our $202.81$~{\rm km} link has a total loss $32.45$~{\rm dB}. This is equivalent to $162.25$~{\rm km} of standard fiber with attenuation of $0.20$~{\rm dB/km}. Besides the longest distance of $202.81$~{\rm km} of fiber link, the experiments with $27.27$~{\rm km}, $49.30$~{\rm km}, $69.53$~{\rm km}, $99.31$~{\rm km}, and $140.52$~{\rm km} of fiber link have also been done to show the performance of the CV-QKD system at different distances. Note, the scheme with the transmitted LO may open loopholes for the eavesdropper to intercept the secret keys by manipulating the intensity of LO pulses~\cite{Ma_PhysRevA_2013,Jouguet_PhysRevA_2013}. To prevent such side-channel attacks, in the experiment, the photodiode for the synchronization is also used for the LO monitoring and we will randomly switch off the signal pulses to do the real time shot-noise unit calibration~\cite{YC_PhysRevApplied_2020}. Thus, the system is immune to most current side-channel attacks against LO. A better way to avoid these attacks is to generate the LO ``locally'' at Bob with a second laser~\cite{Qi_PhysRevX_2015,Soh_PhysRevX_2015}.

\begin{figure}[b]
\centerline{\includegraphics[width=0.54\textwidth]{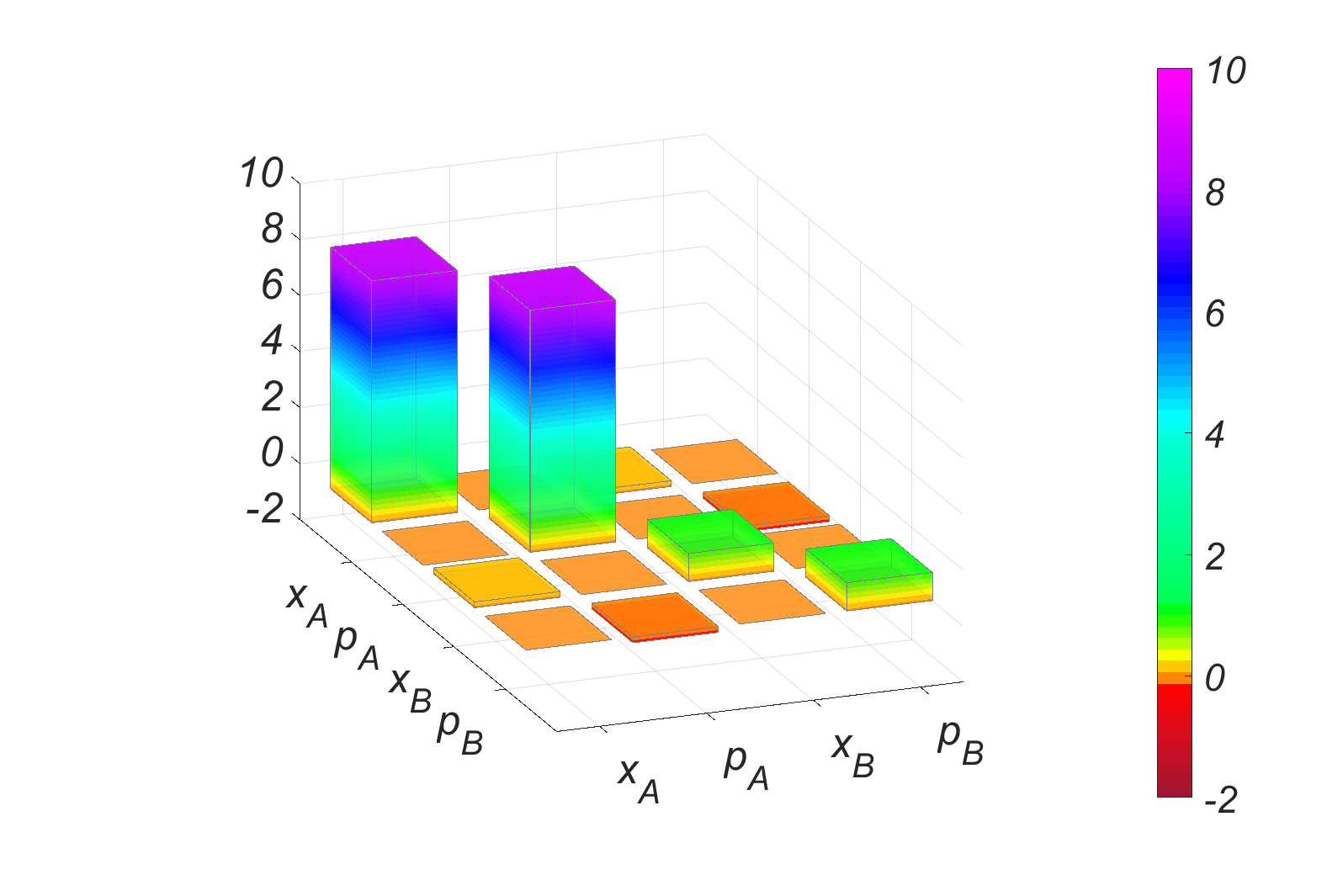}}
\caption{Covariance matrix. We depict the covariance matrix of Alice's and Bob's variables after $202.81$~{\rm km} of fiber link transmission, which is based on the calibrated parameters in TABLE~\ref{table1} using a standard expression. This follows the ordering $\{x_A,p_A,x_B,p_B\}$ and its components are expressed in shot-noise units.}\label{fig2}
\end{figure}

\begin{table*}[t]
	\caption{\label{tab:degree_distribution} Overview of experimental parameters and performances for different fiber lengths. SNR: signal-to-noise ratio; $\beta $: reconciliation efficiency; $\alpha$: system overhead; $FER$: frame error rate of the reconciliation; $V_{A}$: modulation variance; $\xi$: excess noise at the channel input; $\xi'$: worst-case excess noise estimator at the channel input; $\nu_{el}$: electronic noise; $P_{LO}~$: power of LO inside homodyne detector; $\eta$: efficiency of the homodyne detector; $K_{finite}$: final secret key rate in the finite-size regime. {\rm SNU}: shot-noise unit.} 
	\footnotesize\rm
	\begin{ruledtabular}
		\begin{tabular}{c|c c c c c c p{5cm}|}
			Attenuation~({\rm dB}) & 4.36 & 8.29 & 11.68 & 15.89 & 23.46 & 32.45 \\
			\hline
            Length~({\rm km}) & 27.27 &49.30 & 69.53 & 99.31 & 140.52 & 202.81\\
			SNR & 2.8035 & 1.0715 & 0.4619 & 0.1806 & 0.0308 & 0.0023 \\
            $\beta~(\%) $& 95.00 & 95.00 & 96.00 & 96.00 & 96.00 & 98.00 \\
            $\alpha~(\%) $& 10 & 10 & 10 & 10 & 10 & 10 \\
            $FER~(\%) $& 50 & 50 & 10 & 10 & 10 & 90 \\
            $V_{A}$~({\rm SNU}) & 14.37 & 14.14 & 14.12 & 14.53 & 14.23 & 7.65 \\
			$\xi$~({\rm SNU}) & 0.0015 & 0.0033 & 0.0049 & 0.0063 & 0.0086 & 0.0081 \\
            $\xi'$~({\rm SNU}) & 0.0016 & 0.0037 & 0.0058 & 0.0085 & 0.0219 & 0.0383 \\
			$\nu_{el}~$({\rm SNU})& 0.1216 & 0.1881 & 0.2411 & 0.1893 & 0.2717 & 0.1523 \\
            $P_{LO}$~({\rm $\mu$W})& 24.50 & 14.96 & 11.09 & 14.85 & 9.57 & 19.07 \\
			$\eta$~(\%)&61.34 & 61.34 & 61.34 & 61.34 & 61.34 & 61.34 \\
			$K_{finite}$~({\rm bps}) & $2.78\times 10^{5}$ & $0.62\times 10^{5}$ & $4.28\times 10^{4}$ & $1.18\times 10^{4}$ & $318.85$ & $6.214$ \\
		\end{tabular}
	\end{ruledtabular}\label{table1}
\end{table*}

To overcome the channel perturbations due to the long-distance fiber, we employ several automatic feedback systems to calibrate polarization and phase and implement clock and data synchronization. Clock synchronization and data synchronization is implemented by splitting a part of the LO pulses, after their demultiplexing at Bob's side, and detecting them by using a photodiode. The detection results are feed into a clock chip to generate the high-frequency clock signals as the time baseline for the overall system. Combined with the inserted specific training sequences before the data frame for the modulation on $\mathrm{AM}_1$, data synchronization is thus realized using this detection output of the photodiode by distinguishing the training sequences. The polarization calibration module aims to compensate the polarization drift during the transmission through the quantum channel, including the aforementioned polarization beam splitter and photodiode, and another key component, i.e., a dynamic polarization controller. The output of the photodiode is a feedback signal which is used for adjusting the dynamic polarization controller so as to let the LO power larger than that of the signal up to $30$~{\rm dB}. To reduce the excess noise, we use an erbium doped fiber amplifier. As previously mentioned, this amplifies the LO to the optimal working point of the homodyne detector (see the Supplemental Material~\cite{SupplementalMaterial} for more details). The main contribution to the excess noise comes from residual phase noise due to the mismatch between the actual phase noise accumulated in the fiber and its estimation during our process of compensation (see the Supplemental Material~\cite{SupplementalMaterial} for more details).

In our system, we adopt a high-precision phase compensation scheme in order to (i) eliminate the phase shift induced from the unbalanced MZI structure and (ii) decrease the phase noise as much as possible. In this way, we can achieve long-distance transmission where even a slight increase of the excess noise may have nontrivial degradation effects on the value of the secret key rate. To reduce this residual phase noise, we switch the key signals (i.e., those used for key generation) with higher-intensity `reference signals' that are specifically dedicated to phase noise estimation. In our setup, the variance $V_A$ of the Gaussian modulation of the key signals as well as the signal-to-noise ratio (SNR) of the reference signals are controlled by modulators AM$_1$ and AM$_4$, which are in turn controlled by a $10$-bit digital-analog-converter (DAC). In our phase compensation scheme, a series of specific phase reference frames modulate the $\mathrm{AM}_1$ to generate $100$ reference pulses every $1000$ data pulses. With the help of $\mathrm{AM}_4$, the final SNR of the reference pulses is $34$~{\rm dB} higher than that of the signal pulses. Note, sending higher-intensity references pulses here will not increase more practical security problems than sending weak-intensity references pulses~\cite{Qi_PhysRevX_2015}. With this value, the residual phase noise is low enough to support the system over $202.81$~{\rm km} fiber. The results of the homodyne detection are used for calculating the relative phase difference between transmitted and received data, and the phase modulator on the LO path at Bob's side modulates Bob's data according to the calculated phase estimation results.

Once Bob has measured the states sent from Alice, the two parties postprocess their data to generate a secret key~\cite{Zhang_QuantumSciTechnol_2019}. In a CV-QKD system, postprocessing can be divided into four parts: basis sifting, parameter estimation, information reconciliation (or error correction), and privacy amplification. In our system, parameter estimation is performed after error correction (apart from a preliminary estimation of the SNR).

\emph{Results.} $-$ From the correlated data, we compute the covariance matrix illustrated in Fig.~\ref{fig2} from which we can derive the asymptotic key rate of our system. Then taking finite-size effects into account, the reverse-reconciliation secret key rate is given by the general formula~\cite{Leverrier_PhysRevA_2010,Pirandola_RevModPhys_2019}
\begin{equation}
K_{finite} = f\left( {1 - \alpha } \right)\left( {1 - FER} \right)\left[ {\beta I\left( {A:B} \right) - \chi \left( {B:E} \right) - \Delta \left( n \right)} \right],
\end{equation}
where $f$ is the repetition rate, $\alpha$ is the system overhead quantifying the ratio between reference and key signals, FER is the frame error rate of the reconciliation, $\beta$ is the reconciliation efficiency, $I(A:B)$ is the classical mutual information between Alice and Bob, $\chi(B:E)$ bounds Eve's Holevo information on Bob's variable in the finite-size regime, and $\Delta \left( n \right)$ is an offset term which accounts for privacy amplification in the finite-size regime. Here we build corresponding maximum-likelihood estimators for the channel parameters, compute their confidence intervals, and bound their values adopting $6.5$ standard deviations. This worst-case scenario is then used to evaluate Eve's Holevo bound.

\begin{figure}[t]
\centerline{\includegraphics[width=0.52\textwidth]{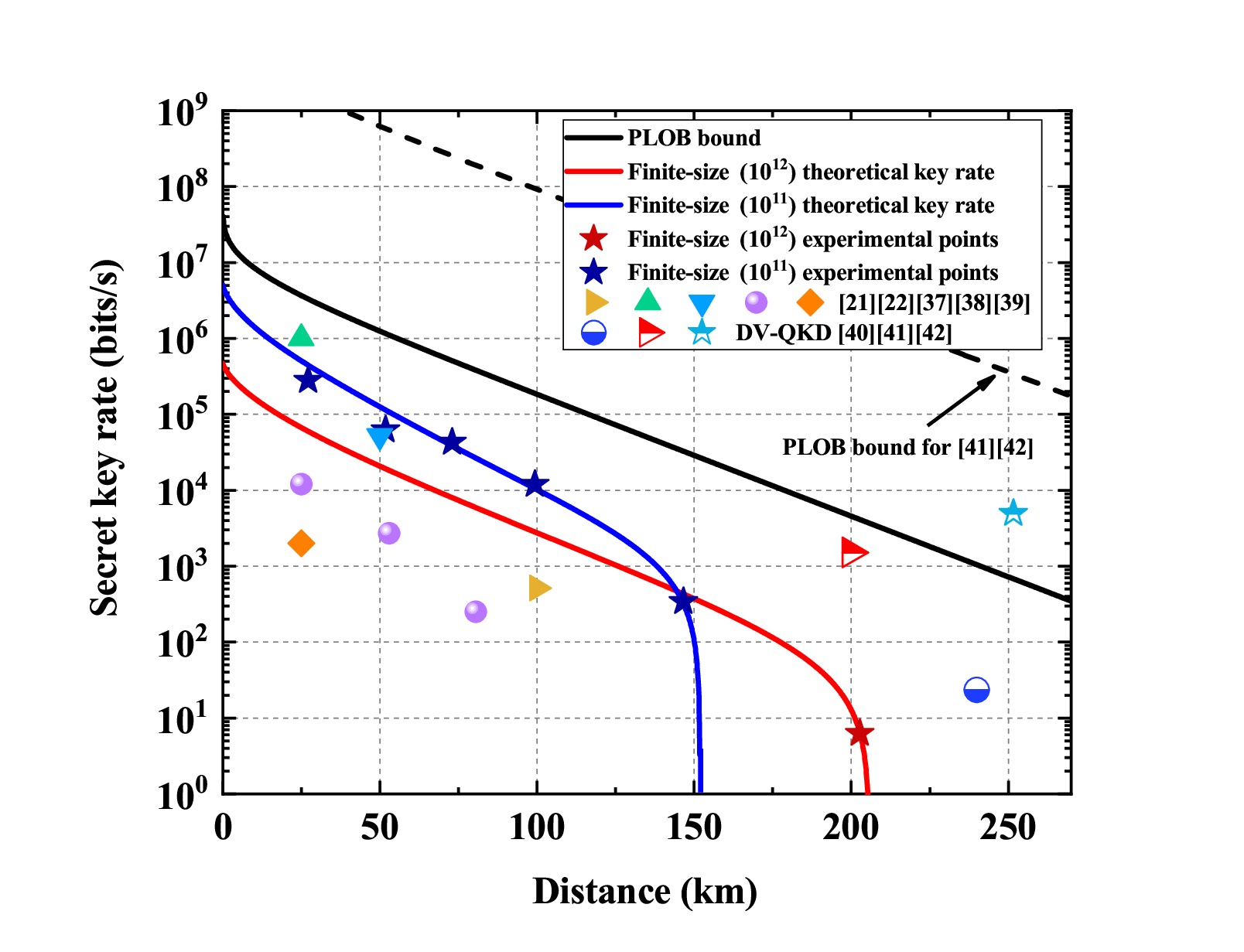}}
\caption{Experimental key rates and numerical simulations. The six five-pointed stars correspond to the experimental results at different fiber lengths of $27.27$~{\rm km}, $49.30$~{\rm km}, $69.53$~{\rm km}, $99.31$~{\rm km}, $140.52$~{\rm km}, and $202.81$~{\rm km}. The blue solid curve is a numerical simulation of the key rate which is computed starting from the experimental parameters at $140.52$~{\rm km}. The red solid curve is the corresponding numerical simulation computed from the parameters at $202.81$~{\rm km}. For comparison, we also show previous state-of-the-art CV-QKD experimental results~\cite{Jouguet_NaturePhoton_2013,Huang_ScientificReports_2016,Huang_OptExpress_2015,Wang_SciRep_2015,Lodewyck_PhysRevA_2007}, DV-QKD experimental results from $200$ to $250$~{\rm km}~\cite{Bernd_Optica_2017,Alberto_ApplPhysLett_2018,Alberto_PhysRevLett_2018}, and we compare the values and the scaling of our rates with the PLOB bound~\cite{Pirandola_NatCom_2017}, i.e., the fundamental limit of repeaterless quantum communications. In particular, the PLOB bound is plotted with respect to our clock rate ($5$~{\rm MHz}, black solid line)
and with respect to the clock rate of Ref.~\cite{Alberto_ApplPhysLett_2018,Alberto_PhysRevLett_2018} ($2.5$~{\rm GHz}, black dashed line).}\label{fig4}
\end{figure}

Highly efficient postprocessing is needed to achieve long transmission distances at sufficiently high secret key rates. Polar codes and multiedge type LDPC codes are used to obtain high reconciliation efficiencies at the intermediate distances (see the Supplemental Material~\cite{SupplementalMaterial} for more details). For extremely low SNRs (lower than $-26$~{\rm dB}), error correction is very difficult and it becomes challenging to construct suitably fixed-rate error correcting codes. For this reason, for our longest distance, we resort to a Raptor code. This is a type of rateless code able to reach high reconciliation efficiency by sending check information until error correction is successful. In order to extract secret key at low SNR (lower than $-26$~{\rm dB}), we use the reconciliation scheme that combine multidimensional reconciliation and Raptor codes. Alice and Bob divide their data into vectors of size $8$ and normalized them. A binary random sequence $c$ is generated by a quantum random number generator. The sequence $c$ is then encoded into another sequence $U$ through Raptor encoding. Bob uses $U$ and his own Gaussian variable $Y'$ to calculate the mapping function $M(Y',U)$. After Alice received enough side information, she starts to recover $U$ by Raptor decoding and, therefore, Bob's random binary sequence $c$. The rate of Raptor codes is uncertain before information transmission. In this work, we set the expected reconciliation efficiency to 98\% and the length of the encoded sequence is $1.82\times 10^{6}$. We achieved high efficient reconciliation error correction with the optimal degree distribution of Raptor code. Then, in the step of privacy amplification, we use hash function (Toeplitz matrices in our scheme) to distill the final key at speed $1.35$~{\rm Gbps} using graphic processing unit after error correction.



The overview of experimental parameters and performance for different fiber lengths is shown in TABLE~\ref{table1}. The secret key rate is $278$~{\rm kbps} at $27.27$~{\rm km} ($4.36$~{\rm dB} losses), $62.00$~{\rm kbps} at $49.3$~{\rm km} ($8.29$~{\rm dB} losses), $42.8$~{\rm kbps} at $69.53$~{\rm km} ($11.68$~{\rm dB} losses), $11.8$~{\rm kbps} at $99.31$~{\rm km} ($15.89$~{\rm dB} losses) and $318.85$~{\rm bps} at $140.52$~{\rm km} ($23.46$~{\rm dB} losses). For the longest transmission distance of $202.81$~{\rm km} ($32.45$~{\rm dB} losses) in our experiment, there is a SNR of $0.0023$, the modulation variance $V_A$ is $7.65$ SNU, and the excess noise is $0.0081$ SNU, where SNU represents shot-noise units. Bob's detector is assumed to be inaccessible to Eve and it is characterized by an electric noise of $0.1523$ SNU and an efficiency of $0.6134$. To obtain enough data, we run the system over seven periods corresponding to a total of $61.73$~{\rm h} of acquisition time (see the Supplemental Material~\cite{SupplementalMaterial} for more details), including the necessary interruptions for alignment. This allowed us to extract $1380889$ secret bits, which corresponds to a secret key rate of $6.214$~{\rm bps}.


The secret key rates of numerical simulations and experimental results are shown in Fig.~\ref{fig4}. The five-pointed stars correspond to our experimental results at different fiber lengths. The blue ones are for $27.27$~{\rm km} and $49.3$~{\rm km} with a reconciliation efficiency of $95$\%, and for $69.53$~{\rm km}, $99.31$~{\rm km}, and $140.52$~{\rm km} with a reconciliation efficiency of $96$\%. The red one is instead for $202.81$~{\rm km} with $98$\% reconciliation efficiency. The blue and red solid curves are the numerical simulations calculated from experimental parameters at $140.52$~{\rm km} and $202.81$~{\rm km} with optimal postprocessing, respectively. We compare our points with the previous state-of-the-art CV-QKD experimental results~\cite{Jouguet_NaturePhoton_2013,Huang_ScientificReports_2016,Huang_OptExpress_2015,Wang_SciRep_2015,Lodewyck_PhysRevA_2007}, DV-QKD experimental results from $200$~{\rm km} to $250$~{\rm km}~\cite{Bernd_Optica_2017,Alberto_ApplPhysLett_2018,Alberto_PhysRevLett_2018}, and we also show their behavior with respect to the PLOB bound~\cite{Pirandola_NatCom_2017}, i.e., the secret key capacity of the bosonic lossy channel and fundamental limit of repeaterless quantum communications. Note that the current record of $502$~{\rm km} in optical fiber has been achieved by the phase-matching DV-QKD of Ref.~\cite{Fang_NaturePhoton_2020}, which is not a one-way protocol but exploits an intermediate twin field node~\cite{Lucamarini_Nature_2018}.

\emph{Conclusion.} $-$ In conclusion, our long distance experiment has extended the security range of a CV-QKD system to the record fiber-distance of $202.81$~{\rm km}, a distance that closes the gap with the performance of current one-way QKD protocols with discrete variable systems. In addition, thanks to optimization procedures at the optical layer (phase compensation) and highly efficient postprocessing techniques, the secret key rates are higher than previous results in CV-QKD at almost all distances. It is worth to remark that these key rates have been achieved with a repetition rate of only 5~{\rm MHz}, therefore much slower than the clocks used in discrete-variable experiments (of the order of $1$~{\rm GHz} or more). While there are still several challenging techniques to achieve high-speed CV-QKD system, such as shot-noise-limited homodyne detector, high-speed postprocessing, and precise parameter estimation, our results pave the way for an implementation of CV-QKD in more practical settings and show that large-scale secure QKD networks are within reach of room-temperature standard telecom components.

We thank Corning Incorporated for providing ultralowloss fiber for the experiment. This work is supported by the Key Program of the National Natural Science Foundation of China under Grant No. 61531003 and the Fund of State Key Laboratory of Information Photonics and Optical Communications.

\smallskip

\begin{widetext}

\section*{Supplementary Material}

\section{Shot-noise-limited high-gain homodyne detector}
Here we show the experimental results for the quantum to classical noise ratio and the bandwidth of the shot-noise-limited high-gain homodyne detector. A 1550-nm fiber-coupled laser (NKT BasiK E15, linewidth 100 Hz) offers continuous wave at its output, which is followed by a variable optical attenuator in order to adjust the beam power to an appropriate value. In the experiment, one of the input ports of beam splitter is left unconnected to provide the vacuum state. The LO and the vacuum state will interfere at the beam splitter with splitting ratio of 50:50. After that, there are a variable optical attenuator and a variable optical delay to balance the output arms of the fiber coupler, so that a homodyne detector will only amplify the differential signal. At the output of the homodyne detector, we then use a spectrum analyzer (Rigol DSA815), an oscilloscope (Keysight Technology MSOS804A) and a FPGA with ADC (ADS5400, samplingrate: 1GHz, sampling accuracy: 12bits) to analyze the output signal in both frequency and time domains. This experimental system needs to be properly adjusted to test the parameters of the homodyne detector.

In the frequency domain, the background noise spectrum of the spectrum analyzer, the electronic noise spectrum of the homodyne detector, and the output noise spectrum of the homodyne detector under different values of the LO power is shown in Fig.~\ref{fig1}. By increasing the LO power, the output noise power of the homodyne detector will rise from kHz to $10$~{\rm MHz}. In the low-frequency region, the lower cut-off frequency is determined by the DC blocking capacitor. However, in this region, the superimposed 1/f noise and the instrument noise are so strong that the output noise spectrum of the homodyne detector is overwhelmed. As illustrated in Fig.~\ref{fig1}, the quantum to classical noise ratio is about $24$~{\rm dB} at an LO power of $0.652$~{\rm mW}.

\begin{figure}[h]
\includegraphics[width=4.8in]{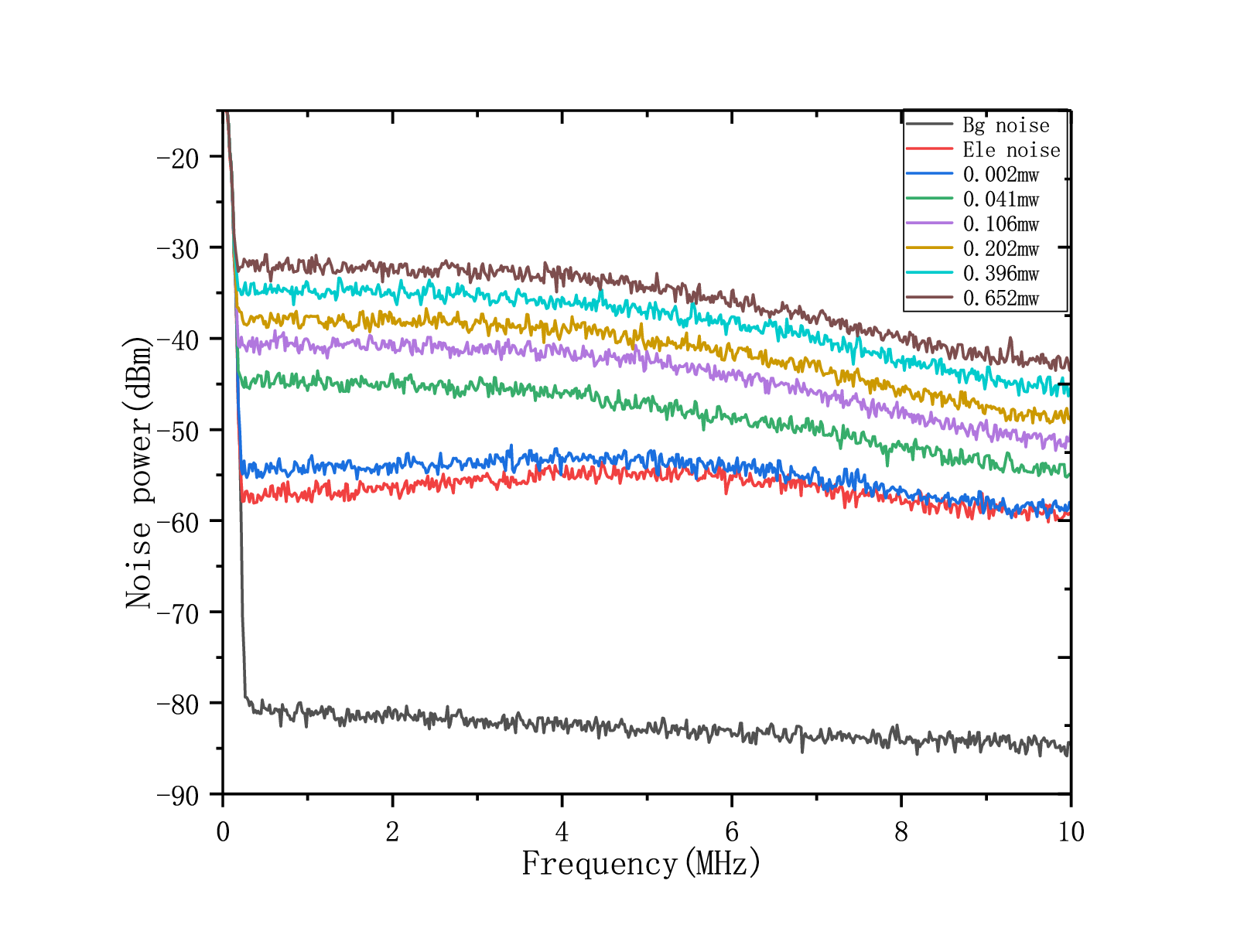}
\caption{Measured noise power of the balanced homodyne detector (BHD) ranges from kHz to $10$~{\rm MHz}. We show that spectrum analyzer background noise spectrum (Bg noise curve), the homodyne detector electronic noise spectrum (Ele noise curve) and the homodyne detector noise spectrum at continuous wave with LO powers of $0.002$, $0.041$, $0.106$, $0.202$, $0.396$ and $0.652$ mW (from the third lowest to the highest curve). Resolution bandwidth: $100$~{\rm kHz}.}
\label{fig1}
\end{figure}

\section{Phase drifting and experimental excess noise results}
Here we explain the problem and management of the residual phase noise in our long distance CV-QKD experiment. In the experiment, Alice draws two sets of Gaussian random variables $\left( {{x_A},{p_A}} \right) \sim N\left( {0,{V_A}} \right)$ and prepares a set of coherent states centered on the point $\left( {{x_A},{p_A}} \right)$. These signal coherent states are transmitted through a fiber, at the end of which Bob performs homodyne detection in order to measure the quadratures of the received signal states (in a switching fashion). Phase noise is a common problem to all coherent detection schemes including those exploited in CV-QKD systems. This is mainly introduced by the unbalanced path of the Mach-Zehnder interferometer (MZI) and the use of different lasers for LO and signal.

\begin{figure}[t]
\includegraphics[width=4.0in]{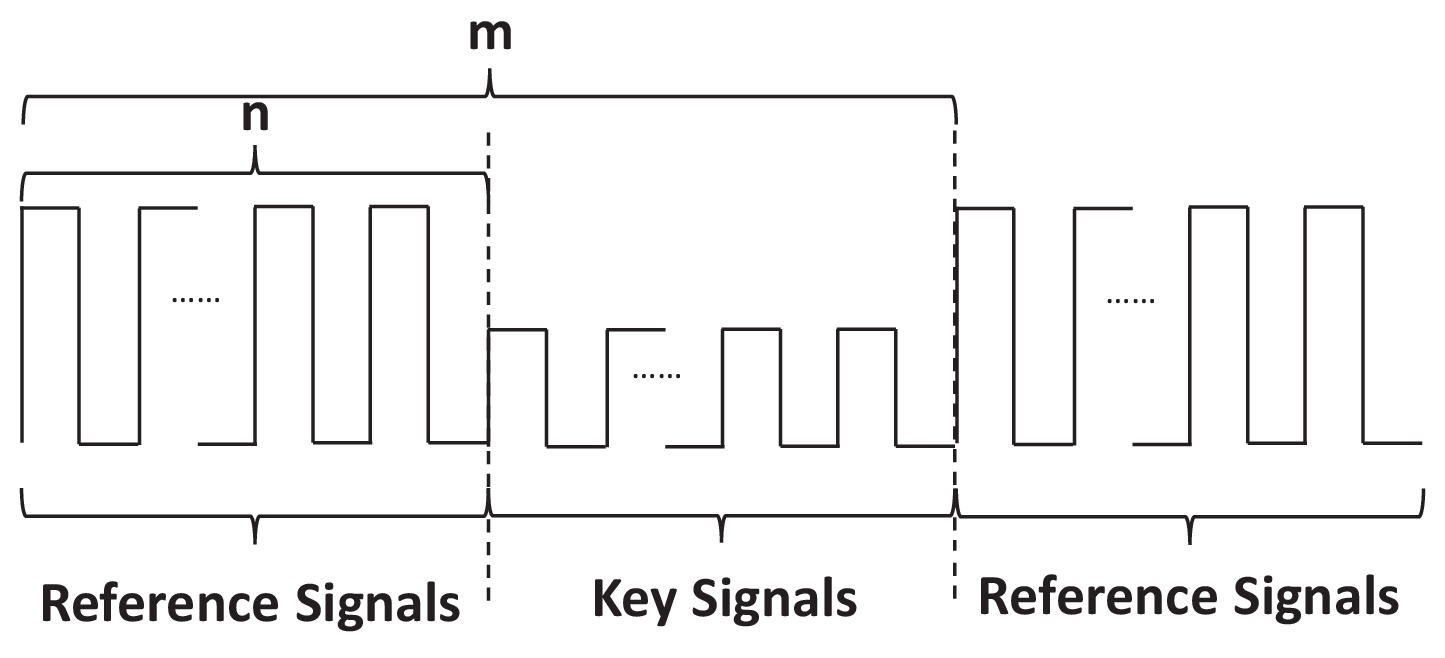}
\caption{The reference signals are used to estimate the phase noise affecting the key signals. The key signals are the Gaussian-modulated coherent states.}
\label{fig2}
\end{figure}

To estimate the phase noise, $n$ reference signals $\left( {{x_A'},{p_A'}} \right)$ are also prepared by Alice and added to the Gaussian-modulated signals or `key signals' for a total compensation period of $m$ signals as shown in Fig.~\ref{fig2}. The ratio $\alpha \equiv n/m$ quantifies the percentage of exchanged signals which are not used for key generation. During a compensation period, the phase noise drift of all measurement results should be small enough to ignore its influence on the secret key rate. The compensation period is dependent on the drift rate of the phase noise and the repetition frequency of system. The phase noise estimation method is the same regardless of whether the drift rate is faster or slower than the repetition frequency.

The measurement results corresponding to the reference signals can be expressed by the formulas~\cite{Qi_PhysRevX_2015,Soh_PhysRevX_2015}
\begin{equation}
\begin{array}{l}
{x_B'} =\eta  \sqrt T \left( {{x_A'}\cos \varphi  + {p_A'}\sin \varphi } \right) + {x_N}\\
{p_B'} =\eta  \sqrt T \left( { - {x_A'}\sin \varphi  + {p_A'}\cos \varphi } \right) + {p_N},
\end{array}
\end{equation}
where $\eta$ is the efficiency of the homodyne detector, $\varphi$ is the phase drift, $T$ is the transmittance of the channel, ${x_N}$ and ${p_N}$ are quadratures describing a mode $N$ with zero mean value and additive-noise variance $V_N$. Here the variance $V_N$ is a global parameter that includes shot noise, excess noise, and the electronic noise of the homodyne detector.  


In the phase noise estimation stage, the measurement results (${x_B'}$ or ${p_B'}$) of the reference signals are transmitted to Alice.  
Since ${x_A'}$, ${p_A'}$, ${x_N}$ and ${p_N}$ are independent of each other, the correlations $E_{XP} = \left\langle {{x_A'}{p_A'}} \right\rangle $, $E_{XN} = \left\langle {{x_A'}{x_N}} \right\rangle $, $E_{PN} = \left\langle {{p_A'}{x_N}} \right\rangle $ are ideally equal to $0$ and $E_{XX} = \left\langle {{x_A'}{x_A'}} \right\rangle $ is equal to $E_{PP} = \left\langle {{p_A'}{p_A'}} \right\rangle $. Ideally, the correlations ${\left\langle {{{p'}_A}{{x'}_B}} \right\rangle }$ and ${\left\langle {{{x'}_A}{{x'}_B}} \right\rangle }$ are given by
\begin{equation}
\begin{array}{l}
\left\langle {{{x'}_A}{{x'}_B}} \right\rangle  = \eta \sqrt T E_{XX}\cos \varphi \\
\left\langle {{{p'}_A}{{x'}_B}} \right\rangle  = \eta \sqrt T E_{PP}\sin \varphi,
\end{array}
\end{equation}
so that the phase drift can be estimated by
\begin{equation}
\tan \varphi = \frac{{\left\langle {{{p'}_A}{{x'}_B}} \right\rangle }}{{\left\langle {{{x'}_A}{{x'}_B}} \right\rangle }}.
\label{eq1}
\end{equation}
Using this value, Alice and Bob can modify their data in order to clean the drift.
However, the estimation of this drift is in practice affected by two errors: the first one is due
to the fact that other types of correlations may intervene in the system (e.g., $E_{XP}$, $E_{XN}$, and $E_PN$
may be non-zero); the second is coming from the fact that the number of reference signals
is finite and therefore there is an intrinsic statistical error due to the finite-size regime.

Because of the difference between the estimated phase drift $\varphi$ and the actual one $\varphi'$, there will
be some phase noise building up in the system and contributing to the overall excess noise. Let us denote this
difference by $\theta = \varphi' -\varphi$ and call it `residual phase noise'. This is now a random variable with
expected value $\kappa  = {\left[ {E\left( {\cos \theta } \right)} \right]^2}$. It is to show that its contributions
to the excess noise is equal to  $\left( {1 - \kappa } \right){V_A}$ where $V_A$ is the variance of the Gaussian modulation of the
input coherent states. Thanks to the compensation system of our experiment, we can keep $\left( {1 - k } \right)$ to low values, which is about $7.6 \times 10^{-5}$

\begin{figure}[t]
\includegraphics[width=3.8in]{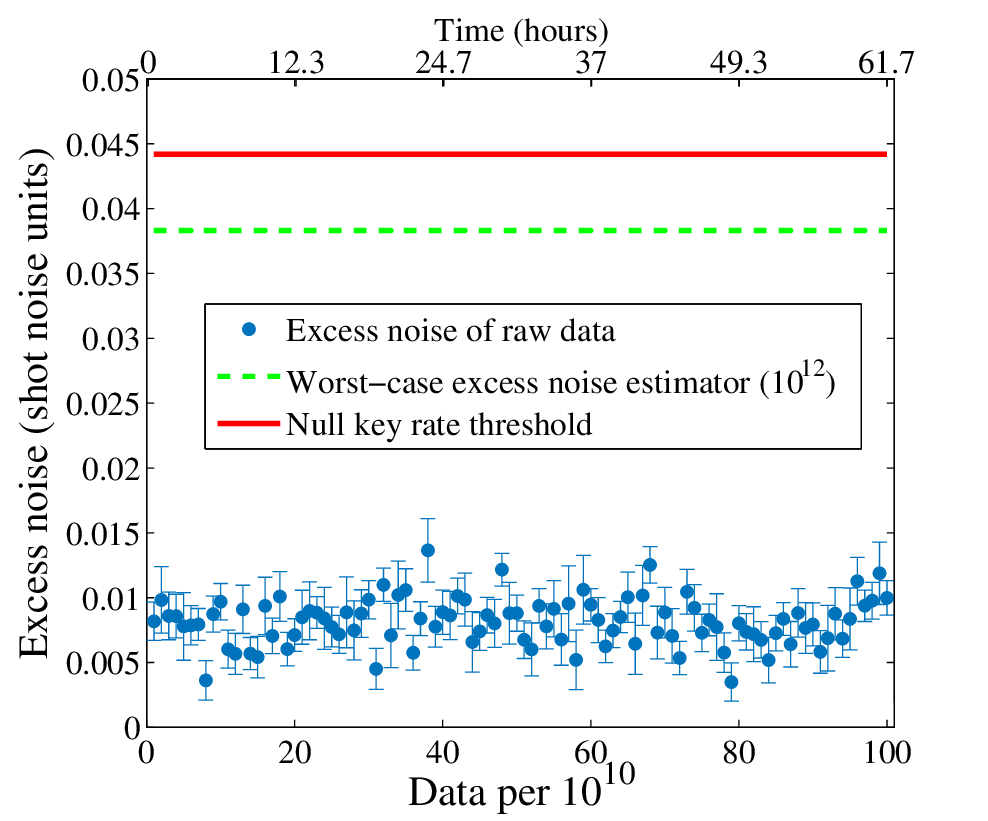}
\caption{Experimental excess noise measured over the data length of $10^{12}$ with a SNR of $0.0023$ for $202.81$~{\rm km} point. The blue plus symbols correspond to measurements performed on blocks of size $1 \times 10^{10}$, each corresponding to roughly $37$ min of data acquisition. The green dash line represents the worst-case excess noise estimator with finite-size effect ($10^{12}$ data length). For comparison, the red line indicates the maximal value of excess noise that allows for a positive secret key rate.}\label{figS3}
\end{figure}

With this value, the residual phase noise is low enough to support the system over $202.81$~{\rm km} fiber. Here we show the experimental results for excess noise measured over the data length of $10^{12}$ with a SNR of $0.0023$ for $202.81$~{\rm km} point. To obtain enough data, we run the system over seven periods corresponding to a total of $61.73$ h of acquisition time, including the necessary interruptions for alignment. As illustrated in Fig.~\ref{figS3}, the worst-case estimator for a block size of $1 \times 10^{12}$ is displayed and is satisfies the condition of extracting secret keys over such long distance.

\begin{figure}[b]
\includegraphics[width=3.8in]{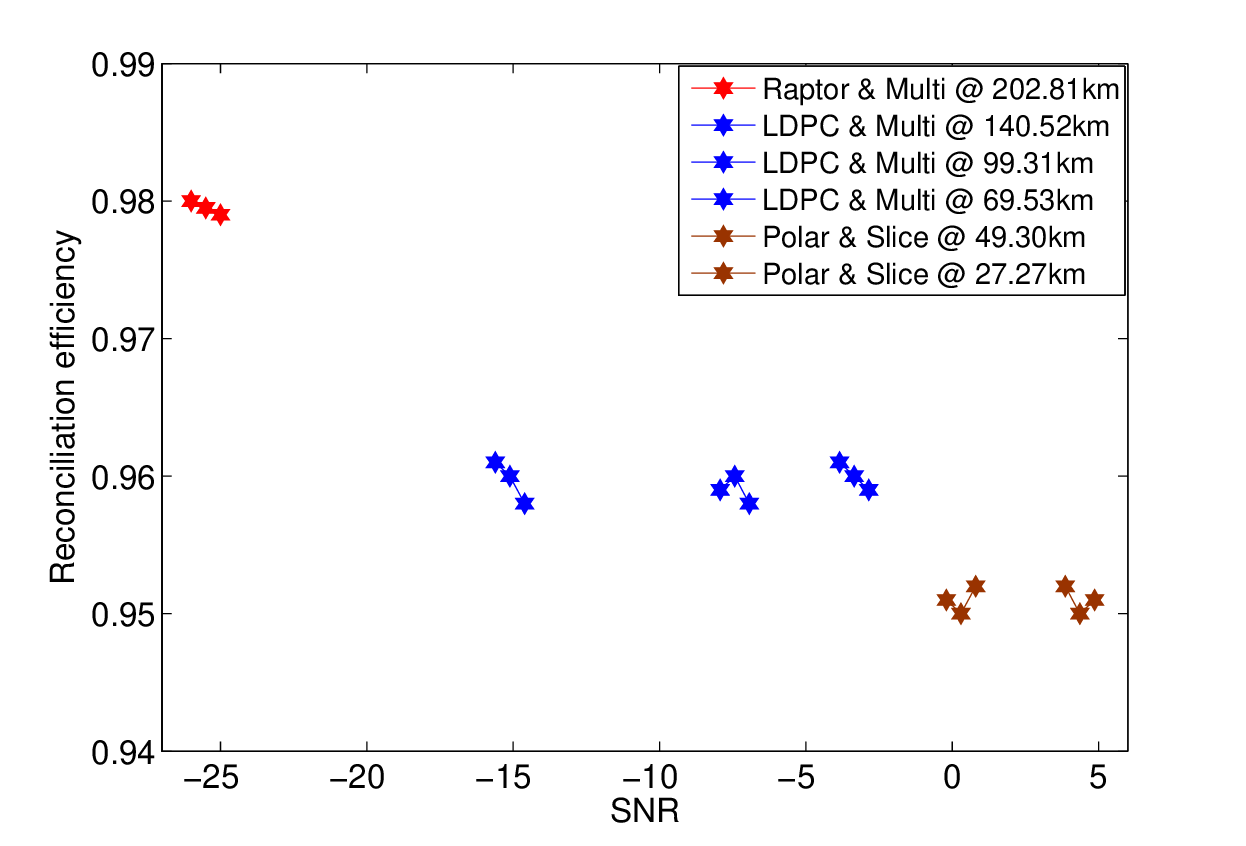}
\caption{Reconciliation efficiencies for different SNRs. We use slice reconciliation with polar codes at $27.27$~{\rm km} and $49.30$~{\rm km}, multidimensional reconciliation with multi-edge type LDPC (MET-LDPC) codes at $69.53$~{\rm km}, $99.31$~{\rm km} and $140.52$~{\rm km}, and multidimensional reconciliation with Raptor codes at $202.81$~{\rm km}.}\label{figS4}
\end{figure}

\section{High efficient information reconciliation at different distances}
Currently, high-performance error reconciliation schemes exists for quantum cryptography~\cite{Zhou_Applied_2019,Wang_QIC_2017,MilicevicThesis,Milicevic17}. In our information reconciliation step, we use slice reconciliation with polar codes at $27.27$~{\rm km} and $49.30$~{\rm km}, multidimensional reconciliation with multi-edge type LDPC (MET-LDPC) codes at $69.53$~{\rm km}, $99.31$~{\rm km} and $140.52$~{\rm km}, and multidimensional reconciliation with Raptor codes at $202.81$~{\rm km}, which is shown in Fig.~\ref{figS4}.

Slice reconciliation scheme is an efficient reconciliation method for the scenario with high SNR. It can be divided in two steps: quantification and error correction. In reverse reconciliation, Bob quantifies his data into $n=5$ slices. Then the real axis is divided in $2^n$ intervals by the interval points which are designed to maximize the quantification efficiency. In particular, we choose equal-width intervals to quantify the Gaussian distribution variables. After the quantification, a multi-level coding (MLC) and multi-stage decoding (MSD) scheme based on polar codes is performed to eliminate the errors. Each level is encoded independently in MLC and decoded successively in MSD with the decoding results used as side information to assist in decoding in the next stage. The optimal code rate of the ${i^{th}}(1 \le i \le n)$ slice with a given SNR is calculated by $R_{opt}^i = 1 - ({I_i}(\infty ) - {I_i}(SNR))$, where ${I_i}(SNR)$ is the mutual information of the ${i^{th}}$ equivalent channel for given SNR~\cite{Assche_TranInfTheory_2004}. Then polar codes with a length of $2^{20}$ are used to correct all the errors between Alice and Bob.

Multidimensional reconciliation scheme transforms a channel with Gaussian modulation to a virtual binary modulation channel as a first step. For 8-dimensional reconciliation, Alice and Bob divide their data into vectors $X$ and $Y$ of size $8$ and normalize them to $X' = \frac{X}{{\left\| X \right\|}}$ and $Y' = \frac{Y}{{\left\| Y \right\|}}$. Then a random binary vector $U$ is generated by quantum random generator. Bob calculates the mapping function $M(Y',U)$ that satisfies $M(Y',U) \cdot Y' = U$. The mapping function is sent to Alice as side information. Alice maps her Gaussian variable $X'$ to $V$ where $V = M(Y',U) \cdot X'$~\cite{Leverrier_PhysRevA_2008}. Then Alice starts to recover $U$ by MET-LDPC codes with a code length of $10^6$. We use three different code rate parity check matrices, one is $0.25$ for $73$~{\rm km}, the other one is $0.1$ for $99.31$~{\rm km} and the last one is $0.02$ for $146.62$~{\rm km}.

\end{widetext}

\end{document}